\documentclass[10pt]{article}
\usepackage{fullpage}

\usepackage{amsmath}
\usepackage{amssymb}
\usepackage{amsfonts}
\usepackage{comment}
\usepackage{color}
\usepackage{cite}
\usepackage{caption}
\usepackage{subcaption}
\usepackage{graphicx}
\usepackage{enumitem}
\usepackage{comment}
\usepackage{bm}
\usepackage{stackrel}
\usepackage{flushend}

\def\##1{{\bf #1}}
\def\=#1{\underline{\underline{#1}}}

\def\+
#1{\underline{\bf #1}}
\def\*#1{\underline{\underline{\bf #1}}}

\def\r#1{(\ref{#1})}

\def\les{\left[}
\def\ris{\right]}
\def\lec{\left\{}
\def\ric{\right\}}
\def\lek{[{\kern 0.1em}}
\def\rik{{\kern 0.1em}]}

\def\.{\mbox{ \tiny{$^\bullet$} }}

\def\eps{\varepsilon}
\def\epso{\eps_{\scriptscriptstyle 0}}

\def\muo{\mu_{\scriptscriptstyle 0}}
\def\etao{\eta_{\scriptscriptstyle 0}}

\def\ko{k_{\scriptscriptstyle 0}}
\def\co{c_{\scriptscriptstyle 0}}

\def\ux{\hat{\#u}_{ x}}
\def\uy{\hat{\#u}_{ y}}
\def\uz{\hat{\#u}_{ z}}

\def\utheta{\hat{\#u}_{{\theta}}}
\def\uphi{\hat{\#u}_{{\phi}}}

\def\ur{\hat{\#{u}}_{ r}}
\def\utheta{\hat{\#{u}}_{ \theta}}
\def\uphi{\hat{\#{u}}_{ \phi}}

\def\kop{k_{{\scriptscriptstyle 0}p}}
\def\fo{\bar{f}_{\scriptscriptstyle 0}}

\def\as{a_{s}}
\def\ap{a_{p}}

\def\ts{t_\mathrm{s}}

\def\smn{_{smn}}
\def\smnp{_{smnp}}

 \def\tQ{{\tilde{Q}}}
 \def\pol{{\bf  {\mathfrak p}}}

\begin{document}

\begin{center}

\LARGE{ {\bf Gaussian pulse scattering by a chiral spherical shell }}
\end{center}
\begin{center}
\vspace{10mm} \large
 
  {H\'ector M. Iga-Buitr\'on}\\
{\em School of Mathematics and
   Maxwell Institute for Mathematical Sciences\\
University of Edinburgh, Edinburgh EH9 3FD, UK}
 \vspace{3mm}\\
 {Tom G. Mackay}\footnote{E--mail: T.Mackay@ed.ac.uk.}\\
{\em School of Mathematics and
   Maxwell Institute for Mathematical Sciences\\
University of Edinburgh, Edinburgh EH9 3FD, UK}\\
and\\
 {\em NanoMM~---~Nanoengineered Metamaterials Group\\ Department of Engineering Science and Mechanics\\
The Pennsylvania State University, University Park, PA 16802--6812,
USA}
 \vspace{3mm}\\
 {Akhlesh  Lakhtakia}\\
 {\em NanoMM~---~Nanoengineered Metamaterials Group\\ Department of Engineering Science and Mechanics\\
The Pennsylvania State University, University Park, PA 16802--6812, USA}\\
and\\
{\em School of Mathematics,
University of Edinburgh, Edinburgh EH9 3FD, UK}

\normalsize

\end{center}

\begin{center}
\vspace{5mm} {\bf Abstract}
\end{center}

Theory was formulated for scattering by a coated chiral sphere of a plane wave of arbitrary polarization state with amplitude modulated by a Gaussian pulse. The spherical core and the concentric shell of the sphere were composed of two different homogeneous materials, both  isotropic chiral. Calculations of energy efficiencies for extinction, total scattering, and absorption were carried out for the shell material with experimentally determined constitutive parameters, the core being vacuous. 
All three energy efficiencies depend on the relative thickness of the shell and the circular polarization state of the carrier plane wave.

\section{Introduction}

Frequency-domain scattering   by a sphere is a classic  problem in electromagnetic field theory. The scattering of a
linearly polarized
  plane wave 
  by a homogeneous sphere composed of an isotropic dielectric material was solved by Lorenz
\cite{Lorenz1890}
 and Mie  \cite{Mie1908} well over a century ago. Subsequently, the theory has been extended to   spheres made of an  isotropic dielectric-magnetic material  \cite{Stratton1941,BH1983} or an isotropic chiral material \cite{Bohren1974,Bhattacharyya1990,Zhen2012}, as well as to concentrically multilayered spheres \cite{Bohren1975,Yung2002,Shang2016,Shang2019}. However, the focus of theoretical work has almost entirely been on linearly polarized incidence. Circularly polarized incidence, which is a more natural choice for
 spheres made of isotropic chiral materials \cite{Beltrami}, and the more general elliptically polarized incidence
  have received scant attention \cite{IgaJOSAA2025,Lak}. 
 {Whereas a few analyses of time-domain scattering   by an isotropic dielectric sphere \cite{Grehan2001,Gouesbet2001,Chrissoulidis2012} 
  (or isotropic dielectric cylinder \cite{Schettini2019})   have been reported, including for a sphere undergoing uniform translational  motion \cite{Garner2017},
   time-domain scattering  by an isotropic chiral sphere has not been  addressed hitherto, to the best of our knowledge.}
 
 In this letter we develop the theory for scattering by a coated chiral sphere of a signal carried by  a plane wave of arbitrary polarization state. A Gaussian pulse is chosen   for amplitude modulation of the carrier plane wave. Numerical results, based on the measured frequency-dependent constitutive parameters
 of a real isotropic chiral material \cite{Gomez2008}, 
 are presented to highlight the dependencies of the energy  efficiencies
for extinction, total scattering, and absorption \cite{Garner2017} on the polarization state of the incident
carrier plane wave.
  
In the  notation is adopted,  vector quantities are denoted by boldface and 
2$\times$2
matrix quantities are enclosed in square brackets and double underlined.
The  unit vectors $\ux$, $\uy$, and $\uz$ relate to the Cartesian coordinate system $(x,y,z)$. The unit vectors $\ur$, $\utheta$, and $\uphi$  are associated with the spherical coordinate system $(r,\theta,\phi)$.
The free-space permittivity, permeability,   wave number, and intrinsic impedance  are denoted by $\epso$,
    $\muo$,   $\ko = \omega \sqrt{\epso \muo}$, and $\etao=\sqrt{\muo/\epso}$, respectively, 
 with $\omega$
being the angular frequency. The speed of light in free space is $\co = 1/\sqrt{\epso \muo}$.
 Also, $i = \sqrt{-1}$ and the complex conjugate is denoted by a superscripted asterisk.
 
\section{Theory} \label{Theory_sec}


\subsection{Coated chiral sphere}

The coated chiral sphere is centered at the coordinate origin $\#r = \#0$.
The core   $r<a$ is made of a homogeneous material labelled 1, whereas the shell  $ a< r < b$ is made of a  homogeneous material labelled 2.
The two materials are specified by the  frequency-domain constitutive relations \cite{Beltrami}
\begin{equation}
\left.
\begin{array}{l}
    \#{D} (\#r,\omega) = \epso\, \eps_\ell (\omega) \, \#{E}(\#r,\omega) + i \, \sqrt{\epso \muo}\, \kappa_\ell 
    (\omega)\, \#{H}(\#r,\omega) 
   \vspace{4pt} \\
    \#{B} (\#r,\omega) =  \muo   \, \mu_\ell(\omega)\, \#{H}(\#r,\omega) - i \, \sqrt{\epso \muo} \, \kappa_\ell(\omega)\, \#{E}(\#r,\omega)
\end{array}
\right\},
\end{equation}
where $\#{E}(\#r,\omega)\in\mathbb{C}^3$, etc., are phasors.
Furthermore,  the relative permittivity $\eps_\ell(\omega)\in\mathbb{C}$, 
relative permeability
 $\mu_\ell(\omega)\in\mathbb{C}$, and  relative chirality parameter $\kappa_\ell(\omega)\in\mathbb{C}$
 for $\ell\in\lec1,2\ric$.

\subsection{Incident signal}
Without loss of generality,
 the incident signal propagates along the $+z$ axis. Thus, the electric field of the incident signal may be expressed  as
\begin{equation}\label{eq_Einc_t1}
\tilde{\bf{E}}_{\mathrm{inc}}(\#r,t)
= 
 \mathrm{Re}\left\{   \pol f(\tau) \right\} ,
\end{equation}
wherein the scalar $\tau = t - ( z/\co)$. 
Following Garner \emph{et al.} \cite{Garner2017},
we set
\begin{equation}\label{eq_pulse}
{
f(\tau) = \exp \left( - i 2 \pi \nu_\mathrm{c} \tau \right) \, \exp\left( - \frac{\tau^2}{2\sigma ^2} \right),
}
\end{equation}
with $\nu_\mathrm{c}$ being the  frequency of the carrier plane wave and $\lambda_\mathrm{c} = \co / \nu_\mathrm{c}$ being the carrier wavelength.
The Gaussian pulse is centered at $t = z/\co$ and the standard deviation $\sigma$ effectively determines the pulse duration.

The  polarization-state vector $\pol = \as \ux + \ap \uy$ with $\as\in\mathbb{C}$ and $\ap\in\mathbb{C}$
specifies an elliptically polarized plane wave propagating along the $+z$ axis. The carrier plane wave
is linearly polarized if either $\as=0$ or $\ap=0$, left-circularly polarized (LCP) if $\ap=i\as$, and
right-circularly polarized (RCP) if $\ap=-i\as$.

For an adequate discrete-time representation of $\tilde{\bf{E}}_{\mathrm{inc}}(\#r,t)$ 
 with sampling period $\ts$, the rate of sampling $1/\ts$ must be at least twice
as great as the highest frequency  in the Fourier transform of the incident signal.
The   total number $N$ of samples has to be selected by trial and error
 to ensure that
 $N \ts$ exceeds the larger of the durations of the incident signal and the scattered signal.
 
The discrete Fourier transform of $f(\tau)$ is given by
\begin{equation}
 \bar{f}(\nu_{p}) = \displaystyle\sum_{q=0}^{N-1} f(\tau_{q}) \exp \left(  i \omega_{p} \tau_{q} \right),
\end{equation}
with
 $\tau_q = q \ts - (z/\co)$, $\omega_p=2\pi\nu_p$, and  $\nu_{p} = p / N \ts$. Hence, inversion of the discrete Fourier transform delivers
\begin{equation} \label{f_tau_q}
 {f}(\tau_{q}) = \dfrac{1}{N} \displaystyle\sum_{p=0}^{N-1} \bar{f}(\nu_{p}) \exp \left( -  i \omega_{p} \tau_{q} \right).
\end{equation}
By substituting Eq.~\r{f_tau_q} into Eq.~\r{eq_Einc_t1}, the
electric field for the incident signal can be approximated by an ensemble of $N$  plane waves as
\begin{equation}\label{eq_Einc_t2}
\tilde{\bf{E}}_{\mathrm{inc}}(\#r,t) \approx   \frac{1}{N} \mathrm{Re} \left\{ \pol \,\displaystyle\sum_{{p=0}}^{{N-1}}   \fo (\nu_{p}) \, \exp \les i \left( \kop z  -  \omega_{p} t \right) \ris \right\},
\end{equation}
where $\fo(\nu_{p})$ denotes the Fourier transform of ${f}(\tau)$ at $z=0$.
Each plane-wave component  has 
 frequency
$\nu_p$,
  wavenumber $\kop = 2 \pi \nu_p / \co$, and 
  electric field phasor 
\begin{equation}\label{eq_Einc1}
    \#{E}_{\mathrm{inc}}(\#r,2\pi\nu_p) = \frac{\fo (\nu_{p})  \, \exp(i \kop z)}{N}\, \pol .
\end{equation}

Following Lorenz--Mie theory \cite{Stratton1941,BH1983},
$  \#{E}_{\mathrm{inc}}(\#r,\omega_p) $ is expressed in terms of
 the vector spherical harmonics 
 $\#{M}\smn^{(1)} \left( \kop \#r \right)$ and $ \#{N}\smn^{(1)} \left( \kop \#r \right) $ as
\begin{equation} \label{eq_Einc_Mie}
\#{E}_{\mathrm{inc}}(\#r,\omega_p)=
\displaystyle\sum_{s\in\{e,o\}}
\displaystyle\sum_{n=1}^{\infty}
\displaystyle\sum_{m=0}^{n}\Big\{
 D_{mn} \les \mathcal{A}\smnp^{(1)} \#{M}\smn^{(1)} \left( \kop \#r \right) + 
    \mathcal{B}\smnp^{(1)} \#{N}\smn^{(1)} \left( \kop \#r\right) \ris\Big\},
\end{equation}
where the coefficients 
\begin{equation}\label{eqn_incidentCoeff}
\left.
\begin{array}{l}
    \mathcal{A}_{emnp}^{(1)} = - 2 n (n + 1) i^n \ap  \fo(\nu_{p}) \delta_{m1}\\
    \mathcal{A}_{omnp}^{(1)} = 2 n (n + 1) i^n \as  \fo(\nu_{p}) \delta_{m1}
\end{array}
\right\}
\end{equation}
with $ \mathcal{B}_{emnp}^{(1)} = - i \mathcal{A}_{omnp}^{(1)}$ and 
$ 
       \mathcal{B}_{omnp}^{(1)} =  i \mathcal{A}_{emnp}^{(1)}$,  $\delta_{mm^\prime}$ is the {Kronecker} delta, and 
\begin{equation}\label{eq_Dmn}
    D_{mn} = (2 - \delta_{m0})\frac{2n+1}{4n (n+1)} \frac{(n-m)!}{(n+m)!}
\end{equation}
is a normalization factor.
Explicit expressions for  $\#{M}\smn^{(1)} \left( \kop \#r \right)$ and $ \#{N}\smn^{(1)} \left( \kop \#r \right) $
are available elsewhere \cite{IgaJOSAA2025,Lak}.

\subsection{Scattered signal}

Each plane-wave component with electric field phasor  $\#{E}_{\mathrm{inc}}(\#r,\omega_p)$  
 gives rise to a 
 scattered field with electric field phasor 
  \begin{equation} \label{eq_Esca_Mie}
\#{E}_{\mathrm{sca}}(\#r,\omega_p)=
\displaystyle\sum_{s\in\{e,o\}}
\displaystyle\sum_{n=1}^{\infty}
\displaystyle\sum_{m=0}^{n}\Big\{
 D_{mn} \les\mathcal{A}\smnp^{(3)} \#{M}\smn^{(3)} \left( \kop \#r \right) + 
    \mathcal{B}\smnp^{(3)} \#{N}\smn^{(3)} \left( \kop \#r\right) \ris\Big\}
\end{equation}
for $r>b$.
Explicit expressions for  $\#{M}\smn^{(3)} \left( \kop \#r \right)$ and $ \#{N}\smn^{(3)} \left( \kop \#r \right) $  are also available elsewhere \cite{IgaJOSAA2025,Lak}. 

Our task now is to derive expressions for the coefficients
$\mathcal{A}\smnp^{(3)}$ and $\mathcal{B}\smnp^{(3)}$. We do so by enforcing
boundary conditions on the surfaces $r=a$ and $r=b$.
Since the coated sphere is made of isotropic chiral materials, we use the {left- and right-handed} vector spherical Beltrami functions \cite{Bohren1975,Beltrami}
\begin{equation}\label{eq_L_R_shell}
\left.
\begin{array}{l}
\#{L}^{(\ell,g)}\smnp ( \#{r}) = \dfrac{1}{\sqrt{2}}\left[ \#{M}\smn^{(g)}(k^{L}_{\ell p} \#{r}) +  \#{N}\smn^{(g)}(k^{L}_{\ell p} \#{r})\right] \vspace{4pt} \\
\#{R}^{(\ell,g)}\smnp ( \#{r}) = \dfrac{1}{\sqrt{2}}\left[ \#{M}\smn^{(g)}(k^{R}_{\ell p} \#{r}) - \#{N}\smn^{(g)}(k^{R}_{\ell p} \#{r})\right]
\end{array}
\right\}, 
\end{equation}
 with $\ell\in\lec 1,2\ric$, $g \in  \lec 1,3 \ric$, and  the wavenumbers
\begin{equation}\label{eq_kchiral}
\left.
\begin{array}{l}
k^L_{\ell p}
 = 
\kop \les
\sqrt{\eps_\ell(\omega_p)} \sqrt{\mu_\ell(\omega_p)} + \kappa_\ell (\omega_p)\ris \vspace{4pt} \\
k^R_{\ell p}
 = 
\kop \les
\sqrt{\eps_\ell(\omega_p)} \sqrt{\mu_\ell(\omega_p)} - \kappa_\ell(\omega_p) \ris 
\end{array}
\right\}\,.
\end{equation}
Consequently the internal electric field phasor for the spectral component of frequency $\nu_p$ 
may expressed as
\begin{equation}\label{eq_E2_Mie}
\#{E}_{{2}}(\#r,\omega_p)= 
\displaystyle\sum _{g\in\{1,3\}} 
\displaystyle\sum_{s\in\{e,o\}}
\displaystyle\sum_{n=1}^{\infty}
\displaystyle\sum_{m=0}^{n}\Big\{
D_{mn} \les \mathcal{C}\smnp^{(g)} \#{L}\smnp^{(2,g)}(\#{r}) + 
\mathcal{D}\smnp^{(g)} \#{R}\smnp^{(2,g)}(\#{r}) \ris\Big\}
\end{equation}
in the shell $b > r >a$ and
\begin{equation}\label{eq_E1_Mie}
\#{E}_{{1}}(\#r,\omega_p) = 
\displaystyle\sum_{s\in\{e,o\}}
\displaystyle\sum_{n=1}^{\infty}
\displaystyle\sum_{m=0}^{n} \Big\{
D_{mn} \les \mathcal{E}\smnp^{(1)} \#{L}\smnp^{(1,1)}(  \#{r})  + 
\mathcal{F}\smnp^{(1)} \#{R}\smnp^{(1,1)}( \#{r}) \ris\Big\}
\end{equation}
in the core  $  r<a$.

The coefficients   
$ \mathcal{C}\smnp^{{(g)}}$ and  $\mathcal{D}\smnp^{{(g)}}$ in Eq.~\r{eq_E2_Mie} as well as the coefficients $ \mathcal{E}\smnp^{{(1)}}$ and $\mathcal{F}\smnp^{{(1)}}$ in Eq.~\r{eq_E1_Mie} are determined by imposing the standard electromagnetic boundary conditions at $r=a$ and $r=b$, and exploiting the orthogonalities of the trigonometric functions and associated Legendre functions \cite{Bohren1975}. 
Thereby, the scattered-field coefficients
are related to the incident-field coefficients as
\begin{equation} \label{eq_ScaCoeff}
\left( \begin{matrix} \mathcal{A}^{(3)}\smnp \\ \mathcal{B}\smnp^{(3)} \end{matrix} \right) = 
 \left( \begin{matrix}
    {c}_{np} & {d}_{np} \\ 
    {d}_{np} & {e}_{np}
\end{matrix}\right)
\. 
\left( \begin{matrix} \mathcal{A}\smnp^{(1)} \\ \mathcal{B}\smnp^{(1)} \end{matrix} \right).
\end{equation}
Explicit expressions for the matrix components ${c}_{np}$, ${d}_{np}$, and ${e}_{np}$ are provided in Appendix~1.
Parenthetically,in the case of a homogeneous isotropic chiral sphere, compact expressions for ${c}_{np}$, ${d}_{np}$, and ${e}_{np}$ are available elsewhere
 \cite{IgaJOSAA2025,Lak}.
 
At sufficiently great distances from  the coated chiral sphere, the electric phasor for each spectral component of the scattered field may be approximated as
\begin{equation}\label{eq_Esca_far}
\#{E}_{\mathrm{sca}}(\#r,\omega_p) \approx \#{F}_{\mathrm{sca},p} (\theta,\phi) \dfrac{\exp(i\, {\kop} r)}{r},
\end{equation}
where the $r$-independent vector
\begin{equation}
{\#{F}}_{\mathrm{sca},p}  (\theta,\phi) = \frac{1}{\kop} \Big[
{L}_{\mathrm{sca},p}(\theta,\phi) \frac{\utheta + i \uphi}{\sqrt{2}} + {R}_{\mathrm{sca},p}(\theta,\phi) \frac{\utheta - i \uphi}{\sqrt{2}} \Big]\,.
\end{equation}
The complex amplitudes of the left- and right-handed  components of $\kop{\#{F}}_{\mathrm{sca},p} (\theta,\phi) $ are
\begin{equation}
\left.
\begin{array}{l}
{L}_{\mathrm{sca},p}(\theta,\phi) = \dfrac{\Theta_{\mathrm{sca},p}(\theta,\phi) - i \Phi_{\mathrm{sca},p}(\theta,\phi)}{\sqrt{2}} \vspace{0.2cm} \\
{R}_{\mathrm{sca},p}(\theta,\phi) = \dfrac{\Theta_{\mathrm{sca},p}(\theta,\phi) + i \Phi_{\mathrm{sca},p}(\theta,\phi)}{\sqrt{2}}
\end{array}
\right\}\,,
\end{equation}
with
\begin{equation}
\Theta_{\mathrm{sca},p}(\theta,\phi) = 
 \displaystyle\sum_{s\in\{e,o\}}
\displaystyle\sum_{n=1}^{\infty}
\displaystyle\sum_{m=0}^{n}
\bigg\{
(-i)^{n} D_{mn} \left[ -i \mathcal{A}\smnp^{(3)} \ f\smn(\theta,\phi) + \mathcal{B}\smnp^{(3)} \ g\smn(\theta,\phi) \right] 
\bigg\}
\end{equation}
and 
\begin{equation}
\Phi_{\mathrm{sca},p}(\theta,\phi) = 
 \displaystyle\sum_{s\in\{e,o\}}
\displaystyle\sum_{n=1}^{\infty}
\displaystyle\sum_{m=0}^{n}
\bigg\{
(-i)^{n} D_{mn} \left[ i \mathcal{A}\smnp^{(3)} \ g\smn (\theta,\phi) + \mathcal{B}\smnp^{(3)} \ f\smn (\theta,\phi) \right]
\bigg\}\,.
\end{equation}
Explicit expressions for the orientation-dependent coefficients $f\smn(\theta,\phi)$ and $g\smn(\theta,\phi)$
are available elsewhere \cite{IgaJOSAA2025,Lak}.

Finally, the electric field   $ \tilde{\bf E}_{\mathrm{sca}}(\#r,t)$ of the scattered signal
  in the far zone emerges as
\begin{equation}
\tilde{\bf E}_{\mathrm{sca}}(\#r,t) \approx \frac{1}{N} \mathrm{Re} \left\{ \displaystyle\sum_{{p=0}}^{{N-1}}  {\#{F}}_{\mathrm{sca},p} (\theta,\phi)  \dfrac{\exp\les i  \left(\kop r-\omega_{p} t\right)\ris}{r} \right\}\,.
\label{eq_Esca_t2}
\end{equation}

\subsection{Energy efficiencies}

The presence of the  coated chiral sphere results in extinction of the incident signal. This extinction is attributable to both scattering by the sphere and absorption by the chiral materials within the sphere. 

In terms of each spectral component  of frequency   $\nu_{p}$ belonging to the incident signal, the  extinction power efficiency is given by \cite{IgaJOSAA2025,Lak}
\begin{equation}\label{Cext}
{Q}_\mathrm{ext} (\nu_{p}) = -\frac{2  }{\kop^2b^2}  {\mathrm{Re}}\left\{ 
    \displaystyle\sum_{n=1}^{\infty} 
    \left[ (2n+1) (c_{np}+e_{np})\right] 
    - 4 \frac{{\mathrm{Im}}\left(\as \ap^{*} \right)}  {|\pol |^2} 
     \displaystyle\sum_{n=1}^{\infty} 
     \left[ (2n+1) d_{np} \right]
     \right\}
\end{equation}
and the total power scattering efficiency by
\begin{eqnarray}
    \nonumber
   &&  \hspace{-7mm}   {Q}_\mathrm{sca} (\nu_{p}) =
    \frac{2 }{\kop^2b^2} \Bigg(
    \sum^{\infty}_{n=1} (2n+1)\left[{|c_{np}|^2 + |e_{np}|^2 + 2 |d_{np}|^2}  \right] \\
    \label{Csca}
    &&  \hspace{-5mm} - 4\frac{\mathrm{Im}(\as \ap^{*})}{|{\pol}|^2} \mathrm{Re} \left\{ \displaystyle\sum^{\infty}_{n=1} \left[ (2n + 1) (c_{np} + e_{np}) d_{np}^{*} \right] \right\}
    \Bigg)\,. 
\end{eqnarray}
 The power absorption efficiency can then be obtained as ${Q}_\mathrm{abs} (\nu_{p}) ={Q}_\mathrm{ext} (\nu_{p}) -{Q}_\mathrm{sca} (\nu_{p})$.

Whereas ${Q}_\mathrm{ext,sca,abs} (\nu_{p})$ formulated in terms of the time-averaged powers of the incident and the scattered waves are appropriate for the scattering of time-harmonic fields \cite{BH1983,VB2005}, analogous quantities for time-dependent fields require temporal integration. Thus, the energy extinction efficiency, total energy scattering efficiency, and the energy absorption efficiency are given as \cite{GarnerPRA}
\begin{equation}
\tQ_\mathrm{ext,sca,abs} = \frac{| \pol |^2 \ts}{2 {\tilde{U}_\mathrm{inc}} \etao}\frac{1}{N}\displaystyle\sum_{{p=0}}^{{N-1}} 
\les{Q}_\mathrm{ext,sca,abs}(\nu_{p}) |    \fo(\nu_{p})|^2\ris\,,
\end{equation}
where
\begin{equation}
{
    {\tilde{U}_\mathrm{inc}} = \frac{1}{\etao} \displaystyle\int_{-\infty}^{\infty} \left| \tilde{\bf{E}}_{\mathrm{inc}}(\#r,t) \right|^2 \mathrm{d}t = {\frac{ |\pol |^2 \sigma \sqrt{\pi}}{2 \etao}}
    }
\end{equation}
is the  energy density of the incident signal.

\section{Numerical results}
\begin{figure}[h!]
\centering
\includegraphics[width=1\linewidth]{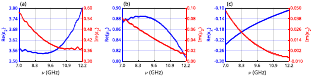}
    \caption{ Real and imaginary parts of {(a)} $\eps_2$,  {(b)} $\mu_2$, and  {(c)} $\kappa_2$   plotted as functions of  
    $\nu$.}
\label{Fig1}
\end{figure}

In order to illustrate the theory developed in \S\ref{Theory_sec} with some numerical results, we set the constitutive parameters $\eps_1 = \mu_1 \equiv1$ and $\kappa_1 \equiv 0$, thereby making the core vacuous \cite{Bohren1975}. For the shell, we chose a composite material for which the constitutive parameters $\eps_2$, $\mu_2$, and $\kappa_2$ had been experimentally determined for  $\nu=\omega/2\pi \in[7.0,12.2]$~GHz   \cite{Gomez2008}. The real and imaginary parts of $\eps_2$, $\mu_2$, and $\kappa_2$ are plotted against    $\nu$ in Fig.~\ref{Fig1}. Clearly, the chosen material 2 is quite dissipative.
 
With $\nu_\mathrm{c}=9.6$~GHz   chosen as the carrier frequency, the standard deviation $\sigma$ of the Gaussian pulse was fixed
equal to $3/\nu_\mathrm{c}$ and the outer radius $b$ of the coated chiral sphere  to
$5 \lambda_\mathrm{c} $. The series in
Eqs.~(\ref{eq_Einc_Mie}), (\ref{eq_Esca_Mie}), (\ref{eq_E2_Mie}), and (\ref{eq_E1_Mie}) were truncated
so that $n\in[1,\bar{N}]$ rather than $n\in[1,\infty)$.
Calculations were then carried out  for increasing values of $\bar{N}$ such that 
the constraints $\Delta_{\rm ext} \leq 1\times 10^{-5}$ and $\Delta_{\rm sca} \leq 1\times 10^{-5}$ were satisfied {for all $\nu_p \in [7,12.2]$GHz},
where
\begin{equation}
\Delta_{\rm ext, sca} = \left|1- \frac{Q_{\rm ext, sca}^{[\bar{N}]}(\nu_p)}{Q_{\rm ext, sca}^{[\bar{N}+1]}(\nu_p)} \right|
\end{equation}
and {$Q_{\rm ext, sca}^{[\bar{N}]}(\nu_p)$} denotes the value of {$Q_{\rm ext, sca}(\nu_p)$} calculated with the truncated series. 
A sample period of $\ts =  0.1/ \nu_\mathrm{c}$ was found to be adequate. Also, convergence was achieved with $\bar{N} \leq 50$, with smaller values of $\bar{N}$   adequate for smaller $b$.

In the interests  of generality, let us present results for an elliptically polarized carrier wave
specified by  
 $\as = 1/\sqrt{5}$ V~m${}^{-1}$ and $\ap = 2i \as$. With the core radius
$a= 4.5 \lambda_\mathrm{c}$, computations were performed for {a duration of $300 \sigma$}.

\begin{figure}[h!]
\centering
\includegraphics[width=1\linewidth]{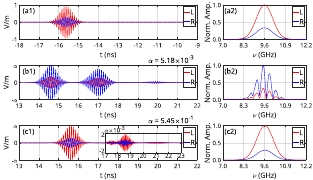}
    \caption{(a1,b1,c1) {Left-handed (L) and right-handed (R)} components of $\tilde{\#E}(\pm30b\uz,t)$
    plotted against $t$, when $\as = 1/\sqrt{5}$ V~m${}^{-1}$ and $\ap = 2i \as$. (a1) Incident signal at $\#r=-30b\uz$, (b1) backscattered
    signal at   $\#r=-30b\uz$, and   
     (c1) forward-scattered signal at   $\#r=30b\uz$.
    (a2,b2,c2) Normalized  Fourier transforms of the left- and right-handed  components of the
    (a2) incident and  (b2) backscattered signals at $\#r=-30b\uz$ and  (c2) the forward-scattered  signal at $\#r=30b\uz$.}
\label{Fig2}
\end{figure}

Figure~\ref{Fig2} presents the time traces of the left- and right-handed components of $\tilde{\bf{E}}_{\mathrm{inc}}(\pm 30b\uz,t)$ vs. $t$. Figure~\ref{Fig2}(a1) isolates the incident signal at $\#r=-30b\uz$,
Fig.~\ref{Fig2}(b1) the backscattered signal at the same location, and Fig.~\ref{Fig2}(c1) the 
forward-scattered signal at $\#r=30b\uz$. The backscattered and the forward-scattered signals contain multiple
pulses arising from back-and-forth pulse propagation within the coated chiral sphere, with the later pulses of diminishing  strength. The first pulse in the forward-scattered signal is stronger than the first pulse in the backscattered signal,
reflecting the higher values of the forward-scattering efficiency compared to the backscattering efficiency
for $\nu\in [7,10]$~GHz. 
The phase difference between the left- and right-handed components observed for the incident signal persist in the    backscattered and the forward-scattered signals.

The normalized Fourier transforms of the time traces in Figs.~\ref{Fig2}(a1,b1,c1) are presented in Figs.~\ref{Fig2}(a2,b2,c2). In each of these three subfigures, the profiles of the normalized Fourier transforms of left- and right handed components are similar.  
The profiles for the forward-scattered signal are very similar to those
of the incident signal, all being either Gaussian (incident) or close to Gaussian (forward-scattered).  
The profiles for the backscattered signal  are very different, comprising at least five distinct peaks fitting inside a Gaussian envelope.

\begin{figure}[h!]
\centering
\includegraphics[width=1\linewidth]{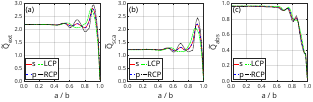}
    \caption{{(a)} $\tQ_\mathrm{ext}$, {(b)} $\tQ_\mathrm{sca}$, and {(c)} $\tQ_\mathrm{abs}$,  plotted against 
    the ratio  $a / b \in (0,1)$ for incident $s$-polarized, $p$-polarized, LCP, and RCP carrier plane waves.}
\label{Fig3}
\end{figure}

Next we turn to the energetics of pulse scattering.
The  energy efficiencies $\tQ_\mathrm{ext,sca,abs}$ are plotted in Fig.~\ref{Fig3} against the ratio  $a / b \in(0,1)$ 
of the core radius to the outer shell radius for the carrier plane wave being: (i)
 $s$ polarized ($\ap=0$), (ii) $p$ polarized ($\as=0$), (iii) LCP ($\ap=i\as$), and (iv) RCP ($\ap=-i\as$).
The plots of $\tQ_\mathrm{ext}$
for $s$ and $p$ polarization states
are indistinguishable from each other, and likewise for  $\tQ_\mathrm{sca}$ and $\tQ_\mathrm{abs}$. In contrast,
the plot of any of the three efficiencies for the LCP state can be distinguished from the plot for the RCP state. 

All three energy efficiencies remain approximately constant
as $a/b$ increases from $0$ to about $0.6$, but variations become evident 
as $a/b$ increases beyond $\sim0.6$ to $1.0$,
regardless of the polarization state of the carrier plane wave.
This behavior can be attributed to the dissipative nature of the shell material.
When the shell is thick ($a / b \lesssim 0.6$), the coated sphere  effectively functions like a homogeneous chiral sphere,  the magnitudes of the pulse fields  penetrating the small vacuous core being negligible.  
As the shell becomes thinner ($a / b \gtrsim 0.6$), variations in the plots
of $\tQ_\mathrm{ext}$, $\tQ_\mathrm{sca}$, and $\tQ_\mathrm{abs}$ become conspicuous 
 since the pulse fields now significantly penetrate the vacuous core. In particular, differences in
 $\tQ_\mathrm{ext}$,  $\tQ_\mathrm{sca}$, and $\tQ_\mathrm{abs}$
  for LCP  and RCP carrier plane waves emerge.
In the limit $(a/b) \to 1$, the effect of the chiral material in the shell    on the incident signal vanishes since the coated sphere becomes entirely vacuous and, accordingly, all three energy efficiencies vanish.

\section{Closing remarks}

We formulated   time-domain scattering   by a coated sphere composed of homogeneous isotropic chiral materials. A Gaussian pulse modulating the amplitude of a carrier plane wave of arbitrary polarization state was adopted as the incident signal. The incident signal was discrete-Fourier-transformed into a finite ensemble of plane waves with different frequencies. For each plane-wave component,
closed-form expressions for the scattered-field coefficients were related to the incident-field
coefficients. While the coated sphere was composed of a spherical core with a concentric shell, the formulation may be straightforwardly extended to a concentrically multilayered sphere. 

The   signal scattered in any direction was obtained by an inverse discrete Fourier transform. The standard extinction, total scattering, and absorption efficiencies for each plane-wave component were used to determine the   energy extinction
efficiency, total energy scattering efficiency, and the energy absorption efficiency of the coated sphere.

A real isotropic chiral material, with experimentally determined constitutive parameters \cite{Gomez2008}, 
was used in numerical studies of a coated sphere with a vacuous core to illustrate the theory.
 Variations in the scattered signal, depending upon the relative thickness of the shell, were studied.
 The energy efficiencies were found to be sensitive to the polarization state of the carrier plane wave when the shell is thin. 
 {The reliability of our results has been confirmed by comparison with results for pulse scattering by an achiral homogeneous sphere (i.e., a sphere for which the relative permittivities 
 $\eps_1 = \eps_2$,
 the relative permeabilities
  $\mu_1 = \mu_2$, and 
 the chirality parameters $\kappa_1 = \kappa_2 = 0$), as considered in Ref.~\cite{Garner2017}.}
 Detailed analyses of diverse numerical results will be reported in due course of time.
 
 In closing, we emphasize that no analysis of time-domain scattering  by an isotropic chiral sphere has  been presented hitherto. Several 
 frequency-domain analyses 
  involving scattering of continuous-wave beams by  isotropic chiral spheres have been undertaken but these analyses are not directly relevant to the present study and their results are not comparable to those presented here.

\section*{Appendix 1}

The 2$\times$2 matrix on the right side of  Eqn.~\eqref{eq_ScaCoeff} may be expressed as 
\begin{align}\label{eq_matrixT} 
& \Bigg( \bigg\{ \left[\=\sigma _{np}^{(1)}(2,b)\right] + \left[\=\sigma _{np}^{(3)}(2,b)\right] \. \left[\=G_{np}\right] \bigg\}^{-1} \. \left[\=\alpha _{np}^{(3)}(b)\right] 
\nonumber
\\
& - \bigg\{ \left[\=\chi _{np}^{(1)}(2,b)\right] + \left[\=\chi _{np}^{(3)}(2,b)\right] \. \left[\=G_{np}\right] \bigg\}^{-1} \. \=\beta _{np}^{(3)}(b) \Bigg)^{-1}  \. 
\nonumber
\\ 
& \Bigg( \bigg\{ \left[ \=\chi _{np}^{(1)}(2,b) \right] + \left[ \=\chi _{np}^{(3)}(2,b) \right] \. \left[ \=G_{np} \right] \bigg\}^{-1} \. \left[ \=\beta _{np}^{(1)}(b) \right] 
\nonumber
\\
& - \bigg\{ \left[ \=\sigma _{np}^{(1)}(2,b) \right] + \left[ \=\sigma _{np}^{(3)}(2,b) \right] \. \left[ \=G_{np}\right] \bigg\}^{-1}  \. \left[ \=\alpha _{np}^{(1)} (b) \right] \Bigg),
\end{align} 
 where the 2$\times$2 matrixes
\begin{eqnarray}\label{eq_matrixG} 
 \nonumber \left[\=G_{np}\right] &=&
 \left\{ \left[\=\sigma_{np}^{(1)}(1,a)\right]^{-1} \. \left[ \=\sigma _{np}^{(3)}(2,a) \right] 
 - \left[\=\chi _{np}^{(1)}(1,a)\right]^{-1}  \. \left[ \=\chi _{np}^{(3)}(2,a) \right] \right\}^{-1} \. 
\\
&& \left\{ \left[\=\chi _{np}^{(1)}(1,a)\right]^{-1} \. \left[ \=\chi _{np}^{(1)}(2,a) \right]
  - \left[\=\sigma_{np}^{(1)}(1,a)\right]^{-1} \. \left[ \=\sigma _{np}^{(1)}(2,a) \right] \right\}
\end{eqnarray}
and
\begin{equation} \label{eq_matrices1} 
\left. 
\begin{array}{l} 
\left[ \=\alpha _{np}^{(g)}(r) \right] = 
	\begin{pmatrix} 
		Z_n^{(g)}(\kop r) & 0 \\ 
		0                 & \zeta_n^{(g)}(\kop r) 
	\end{pmatrix} 
	\vspace{4pt}
\\ 
\left[ \=\beta _{np}^{(g)}(r) \right] = 
	\dfrac{1}{\etao} \begin{pmatrix} 
		\zeta_n^{(g)}(\kop r) & 0 \\ 
		0 & Z_n^{(g)}(\kop r) 
	\end{pmatrix} 
	\vspace{4pt}
\\
\left[ \=\sigma_{np}^{(g)}(\ell, r) \right] = 
	\begin{pmatrix} 
		Z_{n}^{(g)}(k_{\ell p}^L r)     & Z_{n}^{(g)}(k_{\ell p}^R r) \\ 
		\zeta_{n}^{(g)}(k_{\ell p}^L r) & -\zeta_{n}^{(g)}(k_{\ell p}^R r) 
	\end{pmatrix} 
	\vspace{4pt}
\\ 
\left[ \=\chi _{np}^{(g)}(\ell,r) \right] = 
	\dfrac{1}{\eta_\ell} \begin{pmatrix} 
		\zeta_{n}^{(g)}(k_{\ell p}^L r) & \zeta_{n}^{(g)}(k_{\ell p}^R r) \\ 
		Z_{n}^{(g)}(k_{\ell p}^L r)     & - Z_{n}^{(g)}(k_{\ell p}^R r) 
	\end{pmatrix} 
\end{array} 
\right\},
\end{equation} 
for $\ell \in \{1,2\}$ and $g\in\{1,3\}$, 
with $\eta_\ell = \sqrt{\mu_\ell / \eps_\ell}$,
where $\ell=1$ corresponds to the core and $\ell=2$ to the shell. The compact notation
\begin{equation} 
 Z_{n}^{(g)}(w) = \left\{ 
 \begin{array}{ll} j_{n}(w), & g=1 \\ h_{n}^{(1)}(w), & g=3 
 \end{array} 
 \right. 
\end{equation} 
and 
\begin{equation} 
\zeta_n^{(g)}(w) = \dfrac{1}{w} \dfrac{\mathrm{d}}{\mathrm{d}w}\left[ w \ Z_n^{(g)}(w) \right], \qquad g\in\{1,3\}
\end{equation} 
is adopted, with
 $ j_{n}(\cdot)$ being the spherical Bessel function of order $n$ and $ h_{n}^{(1)}(\cdot)$ being the spherical Hankel function of the first kind of order $n$.

%
%

\vspace{3mm}
\noindent
 {\bf Acknowledgments.}
TGM was partially supported  by
EPSRC (grant number APP68512).  AL's research   was partially supported by the Evan Pugh University Professorships Endowment  at Penn State.

\end{document}